# Differential-like Chosen Cipher Attack on A Spatiotemporally Chaotic Cryptosystem

Jiantao Zhou [a*], Wenjiang Pei[a], Jie Huang[a], Aiguo Song[b], Zhenya He[a]

[a]*Department of Radio Engineering, Southeast University, Nanjing, 210096, China*

[b]*Department of Instrument Science & Engineering, Southeast University, Nanjing 210096, China*

## Abstract:

The combinative applications of one-way coupled map lattice (OCML) and some simple algebraic operations have demonstrated to be able to construct the best known chaotic cryptosystem with high practical security, fast encryption speed, and excellent robustness against channel noise. In this paper, we propose a differential-like chosen cipher attack to break the prototypical system cooperating with INT and MOD operations. This cryptographic method exploits the weakness that the high-dimensional cryptosystem degenerates to be one-dimensional under constant driving, therefore, is no more sensitive to the slight perturbation on the driving after convergence. The experimental results show that this method can successfully extract the key within just hundreds of iterations. To the best of our knowledge, it is the first time to present successful cryptanalysis on such a cryptosystem. we also make some suggestions to improve the security in future versions.

## 1. Introduction

Since the pioneering work of Pecora and Carroll, chaos synchronization based cryptography has become a very active research area [1]. However, investigations have shown that a great number of low-dimensional chaotic cryptosystems are fundamentally flawed by a lack of security and various attacks such as nonlinear forecasting, return map, adaptive parameter estimation, error function attack (EFA), and inverse computation based chosen cipher attack, may succeed in recovering messages from the chaotic ciphers [2-6]. Therefore, recent studies have focused on high-dimensional chaotic systems, which have more complex dynamics, higher sensitivity to parameter mismatches and better random properties [7-12]. Particularly, OCML system incorporating with basic algebraic operations such as INT and MOD is found to be able to construct cryptosystems with high practical security [9-12]. S-box algebraic operation and bit-reverse operation are then introduced into such a cryptosystem and reaches higher security with low additional computation expense [10-12]. Furthermore, a two-dimensional map lattice is used for obtaining better efficiency [12]. It has been stated that this class of cryptosystem has optimal overall cryptographic properties, even much better than advanced encryption standard (AES) in security, performance efficiency, and robustness against channel noise, and has been applied to practical voice communication [12-13].

[*] Corresponding author. E-mail address: jtzhou@seu.edu.cn.

Meanwhile, various existing cryptanalysis methods have been implemented to evaluate the security of this class of cryptosystems [2-6]. However, the statistical property based attacks are not effective because of the excellent random statistics of the generated key stream [12]. The parameter estimation based attacks do not work because there are no observable state variables available for constructing convergent cost functions [4,12]. Moreover, one of the most successful attacks, EFA cannot break the system because of two facts [5,12]. Firstly, the error function possesses numerous minima, which leads to an erroneous convergence at one of the local minima. Secondly, the needle-like basin structure precludes effective finding of the tendency towards the secret key by exhaustive searching. Recently, Hu et al. proposed a quite effective chosen cipher attack by using a constant as driving to the receiver [6]. It has been stated that this attack can extract the secret key in prototypical scheme proposed in [9] with relatively much less cost (about $2^{80}$ iterations for four keys) than that of EFA (about $10^{36}$ iterations for four keys) [10]. However, the computational cost is still quite large and no detailed analyses can be found. In this paper, we introduce the differential mechanism into the constant-driving chosen cipher attack, and present a differential-like chosen cipher attack. Under constant driving, the prototypical scheme proposed in [9] degenerates to be one-dimensional, thus after convergence, it will not be sensitive to the slight perturbation on driving in specially selected intervals. It can be shown that the key stream difference before and after the perturbation is a smooth and simple-structured function with respect to the secret key on the whole key space. Therefore, intruder can perturb the driving slightly and observe the key stream difference, and then use a simple optimal searching algorithm to find the key with desired precision only with hundreds of iterations.

This paper is organized as follows. In section 2, we review the spatiotemporal chaos-based cryptosystem incorporating OCML with INT and MOD operations in brief. Differential-like chosen cipher attack together with the experimental results is presented in section 3. Section 4 concludes the paper and some suggestions for further improving the security are outlined.

## 2. OCML based spatiotemporally chaotic cryptosystem

The cryptosystem proposed in [9] uses OCML as the core operation, with the additional INT and MOD operations for keeping robustness of communication against channel noise, and greatly enhancing the sensitivity of chaos synchronization to parameter mismatch, respectively.

Decryption transformation:

$$y_{n+1}(j) = (1-\varepsilon) f_j[y_n(j)] + \varepsilon f_j[y_n(j-1)], \quad f_j(x) = (3.75 + b_j/4)x(1-x), \quad j = 1,2,...,m \tag{1a}$$

$$y_{n+1}(j) = (1-\varepsilon) f[y_n(j)] + \varepsilon f[y_n(j-1)], \quad f(x) = 4x(1-x), \quad j = m+1,...,N \tag{1b}$$

$$K_n^{'} = \mathrm{int}(y_n(N) \times 2^{\mu}) \bmod 2^{\upsilon}, I_n^{'} = (S_n - K_n^{'}) \bmod 2^{\upsilon} \tag{1c}$$

$$y_n(0) = S_n / 2^{\upsilon} \tag{1d}$$

where $K_n^{'}$, $I_n^{'}$, and $S_n$ denote the key stream, the recovered plaintext, and the cipher, respectively. The control parameters $\mathbf{b} = (b_1, b_2 \cdots b_m)$ serving as the secret keys can be wildly chosen in the interval $(0,1)$. The other parameters such as coupling strength, system size etc., which can be opened to intruder, are typically set as $\varepsilon = 0.99$, $\mu = 52$, $\upsilon = 32$, $m = 4$, $N = 25$.

The dynamic forms of the transmitter are exactly the same as those of the receiver with $y_n(j)$, $K_n^{'}$, $I_n^{'}$, and $\mathbf{b} = (b_1, b_2 \cdots b_m)$ replaced by $x_n(j)$, $K_n$, $I_n$, and $\mathbf{a} = (a_1, a_2 \cdots a_m)$, respectively. As the receiver is driven by the transmitted signal as $y_n(0) = x_n(0) = S_n / 2^{v}$, when $\mathbf{b} = \mathbf{a}$, the receiver can reach chaos synchronization with the transmitter, and successfully recover the true plaintext as $y_n(N) = x_n(N)$, $K_n' = K_n$, $I_n' = I_n$.

With the requirement of public-structure and chosen-ciphertext, EFA can be used for investigating the system's key sensitivity [5].

$$e(b_1) = \frac{1}{2^{v} T} \sum_{n=1}^{T} \left| I_n' - I_n \right| \tag{2}$$

where $I_n'$ can be computed by intruder from a fake receiver with designed ciphertext $S_n$ and test key $b_1$. Basin structure of $e(b_1)$ vs $b_1$ is shown in Fig.1. [10] has found that the basin width can be empirically calculated as $W \propto T^{0.3}$, where $T$ is the total number of observed plaintext pairs for each test. As away from the basin, $e(b_1)$ is flat with certain fluctuations, the minimal computational cost for finding four secret keys can be roughly calculated as $\frac{1}{W} \approx 10^{36}$ iterations.

It should be noted that, in the OCML system without INT and MOD operations, large coupling size can also ensure satisfactory confusion and diffusion, but it also prolongs the synchronization time and intensifies the error propagation effect. But with the INT and MOD operation, it is easy to obtain equivalent confusion and diffusion by coupling much less number of lattices, and more important, the resistance against various attacks can be greatly enhanced. In fact, several attacks such as parameter estimation attack, inverse computation based chosen cipher attack can be used to break OCML system without INT and MOD operations, but these methods will fail when with such algebraic operations [4,6,12]. The most important reason is that the state of the last lattice is masked and $2^{\mu-\upsilon}$ computations are needed to derive it from the available key stream $K_n$. For parameter estimation, it is impossible to construct a convergent cost function. While for inverse computation, $2^{\mu-\upsilon}$ computations significantly increase the number of equations for deriving the secret key, and large coupling size further enhances such difficulties. It will be shown later that even with currently strongest chosen cipher attack proposed by Hu et al., the minimal computation cost will reach the level of $2^{80}$ for four parameters [6].

## 3. Differential-like chosen cipher attack

It is well known that chosen cipher attacks are among the strongest attacks if we classify the intensity of various evaluations. The intruder can obtain temporary access to the decryption machinery, which enables him to choose a cipher string and construct the corresponding plaintext. Recently, Hu et al. suggested an effective chosen cipher attack method to evaluate various chaos synchronization based systems by using a constant as the driving [6]. Under such constant driving, the response system will asymptotically converge to certain fixed point related to the driving, thus, the secret key can be recovered by using inverse computations [6]. In fact, the convergent property also holds for the OCML system with INT and MOD algebraic operations. When driven by a constant, for the first lattice,

$$\left|y_{n+1}(1)-y_n(1)\right|=(1-\varepsilon)\left|f_1[y_n(1)]-f_1[y_{n-1}(1)]\right|<4(1-\varepsilon)\left|y_n(1)-y_{n-1}(1)\right| \tag{3}$$

As $0<4(1-\varepsilon)\ll 1$, we have $\lim_{n\to\infty}\left|y_{n+1}(1)-y_n(1)\right|=0$. Repeating the similar computations, we can successively prove: $\lim_{n\to\infty}\left|y_{n+1}(i)-y_n(i)\right|=0 \quad i=2,3...N$. Hence all sites will converge to certain fix points when the driving signal is a constant, irrespective how large the coupling size is.

However such inverse computation based attack is not computationally feasible to break the system with INT and MOD operations, mainly because of the $2^{\mu-\upsilon}$ multi-solutions from the available key stream $K_n$ to $y_n(N)$, making the number of equations to derive the secret key extremely large. Moreover, large coupling size further increases the complexity of resolving such equations. Since the computational cost added by the coupling size is neglectable compared with that induced by MOD operation, the total cost can be roughly estimated as $2^{m(\mu-\upsilon)}=2^{80}$, obviously very large but still much less than that of EFA $10^{36}$.

The inverse computation based chosen cipher attack does not fully make use of the properties under constant driving. In fact, it can be proved that the previous high-dimensional system will become one-dimensional after convergence, hence no more extremely sensitive to small perturbation on the driving. In the following section, for simplicity while without loosing generality, we take the most sensitive parameter $b_1$ in the first right term of the first state equation as the secret key to explain the differential-like chosen cipher attack. The other nonlinear parameters including $b_1$ in the second term of the first state equation are all set to be public constant 1. Because the cipher can be arbitrarily chosen, we take $y_d=4y_n(0)(1-y_n(0))$ as the equivalent driving of the system. As after convergence all the state variables are constant, from Eq. (1) we can directly derive the converged state of all sites as:

$$y_1 = \frac{R' + \sqrt{R'^2 + 4\varepsilon(R'+1)y_d}}{2(R'+1)} = \varphi(y_d)\,,\quad R' = (3.75 + b_1/4)(1-\varepsilon) - 1 \tag{4a}$$

$$y_j = \frac{R + \sqrt{R^2 + 16\varepsilon(R+1)y_{j-1}(1-y_{j-1})}}{2(R+1)} = \phi(y_{j-1})\,,\ R = 4(1-\varepsilon) - 1 \tag{4b}$$

Consequently, $y_2 = \phi(y_1) = \phi[\varphi(y_d)]\,,\ y_3 = \phi^2[\varphi(y_d)]\,,\ \ldots\,,\ y_N = \phi^{N-1}[\varphi(y_d)]$ (5)

It is clear to find that the previous high-dimensional system becomes one-dimensional, no matter how complex the function $\phi^{N-1}[\varphi(x)]$ may be. From Fig. 2, we also see that there exist some narrow intervals around 0.495, in which the function $\phi^{N-1}[\varphi(x)]$ is slowly changing. It is an indication that when the driving is chosen in these intervals, the system is easier to handle.

On the other hand, although according to Eq. (1c) there are $2^{\mu-\upsilon} = 2^{20}$ possible solutions from $K_N$ to $y_n(N)$, the difference between two variables, say $y_n(N)$ and $y_n'(N)$ can be uniquely determined by their corresponding key stream $K_n'$ and $K_n''$, provided that $\left|y_n'(N) - y_n(N)\right| < 2^{-20}$, as Eq. (6) shown. The observable difference will be shown later of great significance in finding the secret key.

$$y_n'(N) - y_n(N) = (K'' - K')/2^{\mu} \tag{6}$$

When a constant $y_d$ is used as the driving to make the receiver converge, we take $K_1 = \mathrm{int}(y_N \times 2^{\mu}) \bmod 2^{\upsilon}$ as the converged key stream. After convergence, a slight perturbation is added to $y_d$ at time *t*, i.e., $y_d' = y_d + \tau$, $\tau \ll 1$. At time $t+N$ a changed key stream $K_2 = \mathrm{int}(y_N' \times 2^{\mu}) \bmod 2^{\upsilon}$ can be observed. The function $g(b, y_d) = K_2 - K_1$ is defined as *Instant Perturbation Function* (*IPF*), which describes the dependence between the parameter and the key stream difference before and after the perturbation. The perturbation $\tau = 2^{-36}$ if not specified, which ensures that under certain constant driving $y_d$, $\left|g(b, y_d)\right| < 2^{-20}$ holds for any $b \in (0,1)$.

Furthermore, we can prove that, for every secret key $b_1 \in (0,1)$, there exists at least one driving $y_{d1}$ such that $g(b_1, y_{d1}) = 0$, i.e., no difference can be observed before and after the perturbation. We call the zone consisting of such points $(b_1, y_{d1})$ the silent position. The state equations after the perturbation are shown in Eq. (7). Note that all the first terms remain unchanged because the perturbation takes effect to them from the next time.

$$y_1^{'} = (1-\varepsilon)(3.75 + b_1/4) y_1 (1-y_1) + \varepsilon(y_{d1} + \tau) \tag{7a}$$

$$y_j^{'} = 4(1-\varepsilon) y_j (1-y_j) + 4\varepsilon y_{j-1}^{'} (1-y_{j-1}^{'}), \quad j = 2,3,...,N \tag{7b}$$

For $e_k = y_k^{'} - y_k$, from Eq. (1) and (7) we have the difference dynamics:

$$e_1 = \varepsilon\tau, \quad e_j = 4\varepsilon e_{j-1}(1-2y_{j-1}) - 4\varepsilon(e_{j-1})^2, \ j = 2,3,...,N \tag{8}$$

As $\tau \ll 1$, $(e_{j-1})^2 = 0$ in practical computation. Thus, Eq. (8) becomes

$$e_j = 4\varepsilon e_{j-1}(1-2y_{j-1}), \ j = 2,3,...,N \tag{9}$$

Consequently, we have:

$$e_j = \varepsilon\tau(4\varepsilon)^{j-1} \prod_{i=1}^{j-1} (1-2y_i), \ j = 2,3,...,N \tag{10}$$

Hence, any $y_j = 0.5$, $j = 1,2,...,N-1$ leads to $g(b_1, y_{d1}) = 0$. Assuming $y_1 = 0.5$, from Eq. (1) we have: $y_{d1} = [2 - (1-\varepsilon)(3.75 + b_1/4)]/4\varepsilon$. When $b_1$ is freely chosen in $(0,1)$, the range of $y_{d1}$ can be calculated as $\Phi = (0.49494949495, 0.49558080808)$, which is around 0.495 (that's why we choose the interval around 0.495 in Fig. 2). Therefore, for any $b_1 \in (0,1)$ we can find at least one $y_{d1}$ in $\Phi$, satisfying $g(b_1, y_{d1}) = 0$. In Fig. 3 we plot the three-dimensional figure of $|g(b, y_d)|/2^u$ in terms of the secret key chosen in $(0,1)$ and the driving in $\Phi$. The bottom of the mesh represents the silent position. It can be clearly seen that there is at least one $y_{d1}$ for every secret key corresponding to the silent state, which experimentally proves our above conclusion.

Since $\phi^{N-1}[\varphi(x)]$ is slowly changing in $\Phi$, as Fig. 2 shown, a simple optimal searching algorithm shown in Eq. (11) can be used to search one appropriate $y_{d1}$ in $\Phi$.

$$y_{d1}(k+1) = y_{d1}(k) + \gamma(K_2 - K_1)/2^{\upsilon} \tag{11}$$

Moreover, if there is one point $(b_1, y_{d1})$ satisfying $g(b_1, y_{d1}) = 0$, then there exists one line $Y = AX + B$ containing the point $(b_1, y_{d1})$, on which any point, say $(b^*, y^*)$, satisfies $g(b^*, y^*) = 0$. If $b_1$ and $y_{d1}$ are used as the secret key and the constant driving, respectively, the converged state of the first lattice is $y_1 = \varphi(b_1, y_{d1})$, shown in Eq. (4a). From Eq. (5) and (10), we know that $g(b_1, y_{d1})$ is uniquely determined by $y_1$. Hence, any point $(b^*, y^*)$ making $y_1|_{b=b^*, y_d=y^*} = \varphi(b_1, y_{d1})$ leads to $g(b^*, y^*) = g(b_1, y_{d1}) = 0$. Then from Eq. (4a), we can obtain the line equation:

$$y^* = \varphi(b_1, y_{d1})/\varepsilon + \varphi(b_1, y_{d1})(1-y_1)(1-\varepsilon)(3.75 + b^*)/\varepsilon \tag{12}$$

Therefore, there is a linear dependence between the secret key and the driving corresponding to the silent state. As a result, the task of finding the secret key can be converted to find the corresponding driving. We can use the iterative method described in Eq. (11) with the middle point of $\Phi$ as the initial value to search the corresponding driving in $\Phi$ for every test key in $(0,1)$. As Fig. 4 shown, all the possible points $(b^*, y^*)$ belong to five lines. We can experimentally calculate the equations of line 1-5, results averaging over 100 realizations shown in Table 1.

Therefore, the procedure of finding the secret key by using differential-like chosen cipher attack can be as follows (taking a randomly generated number $b_1 = 0.7865302612$ for example).

*Step* 1: Search a constant driving for the authorized receiver using Eq. (11). The step size $\gamma = 0.0001$. After about 100 iterations we can obtain $y_{d1} = 0.495206963987366133$.

*Step* 2: Analytically calculate the possible secret keys by inserting $y_{d1}$ to the line equations. Five corresponding solutions can be obtained, -0.2283274311, 0.3978096430, 0.5934578440, 0.7865302612, and 1.412698303. Obviously one of the points in {0.3978096430, 0.5934578440, 0.7865302612} is the secret key because only these three points are in $(0,1)$. Hence, if we construct three fake receivers using these three keys, we can make sure that one of them will correctly recover the plaintext, shown in Fig. 5.

## 4. Conclusion

The OCML based cryptosystem incorporating with algebraic operations has excellent overall cryptographic properties. This class of cryptosystem has been shown to be resistible against all the existing attack strategies. In this paper, we present a differential-like chosen cipher attack to break the prototypical scheme of this class of system cooperating OCML with INT and MOD operations. The most important factor leading to the success is that the high-dimensional system degenerates into one-dimensional when driven by a constant, thus no more sensitive to slight perturbations on the driving. The numerical experiments results also have demonstrated the capability.

Moreover, the overall security of this class of cryptosystem can be highly improved by using deterministic randomness and its realizable asymptotic model called Lissajious map [14-15]. Actually, similar concept has also been used to enhance chaoticity of spatiotemporal chaos by Li et al. recently [16]. Another spatiotemporal chaos based cryptosystem built on Lissajious map has already been designed. Our current results show that for the modified cryptosystem, the parameter sensitivity can reach the computational precision $2^{-52}$, when the coupling size $N$ is only three, and parallel encryption is feasible even in one-dimensional structure. More important, the improved cryptosystem can effectively resist the differential-like chosen cipher attack because the perturbation function becomes quite complex. The detailed analyses will be given in our forthcoming paper.

## Acknowledgement

This work was supported by the Natural Science Foundation of China under Grant 60133010

and 60102011, the National High Technology Project of China under Grant 2002AA143010 and 2003AA143040, and the Excellent Young Teachers Program of Southeast University.

Captions of Figures

FIG. 1. $T = 1000$. (a) Basin structure of error function, where point $x_0$ in x-axis represents that $|b_1 - a_1| = 2^{-55+|x_0|}$ for $x_0 \neq 0$ ,while $b_1 = a_1$ for $x_0 = 0$. (b) Fluctuation of error function.

FIG. 2. Converged state with respect to the driving around 0.495. (a) $N = 10$ (b) $N = 20$ (c) $N = 24$ (d) $N = 25$.

FIG. 3. $|g(b, y_d)|/2^u$ with respect to the secret key $b \in (0,1)$ and the driving $y_d \in \Phi$.

FIG.4. (a) $b = 0.7865302612$. The procedure of searching $y_d$ using the adaptive method shown in Eq. (11). (b) Driving $y_d$ as a function of the test key $b$ satisfying $g(b, y_d) = 0$. From the left bottom to the right top, the line index is, respectively, 1 to 5.

FIG. 5. Plaintext recovery. The first 1000 data have been discarded. (a) The plaintext (b) The cipher (c) The recovered plaintext by using $b_1 = 0.3978096430$ (d) The recovered plaintext by using $b_1 = 0.5934578440$ (e) The recovered plaintext by using $b_1 = 0.7865302612$.

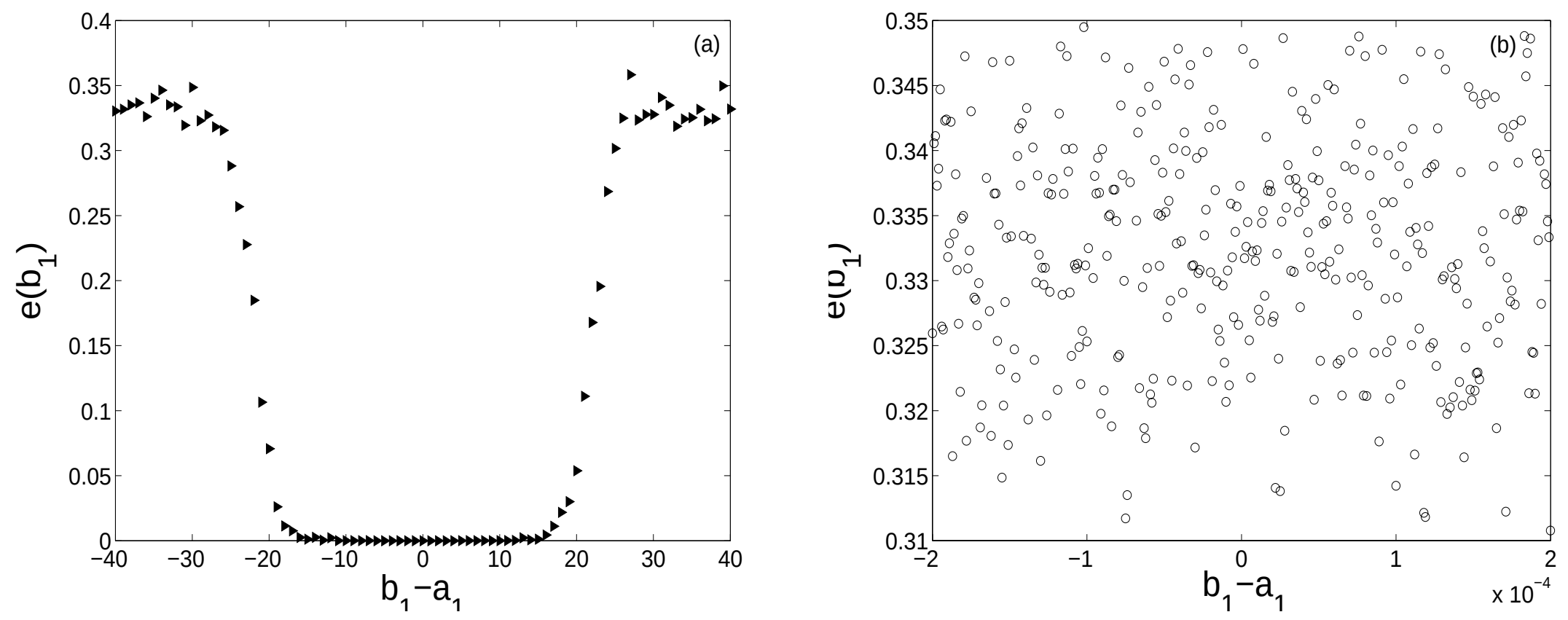

Fig. 1

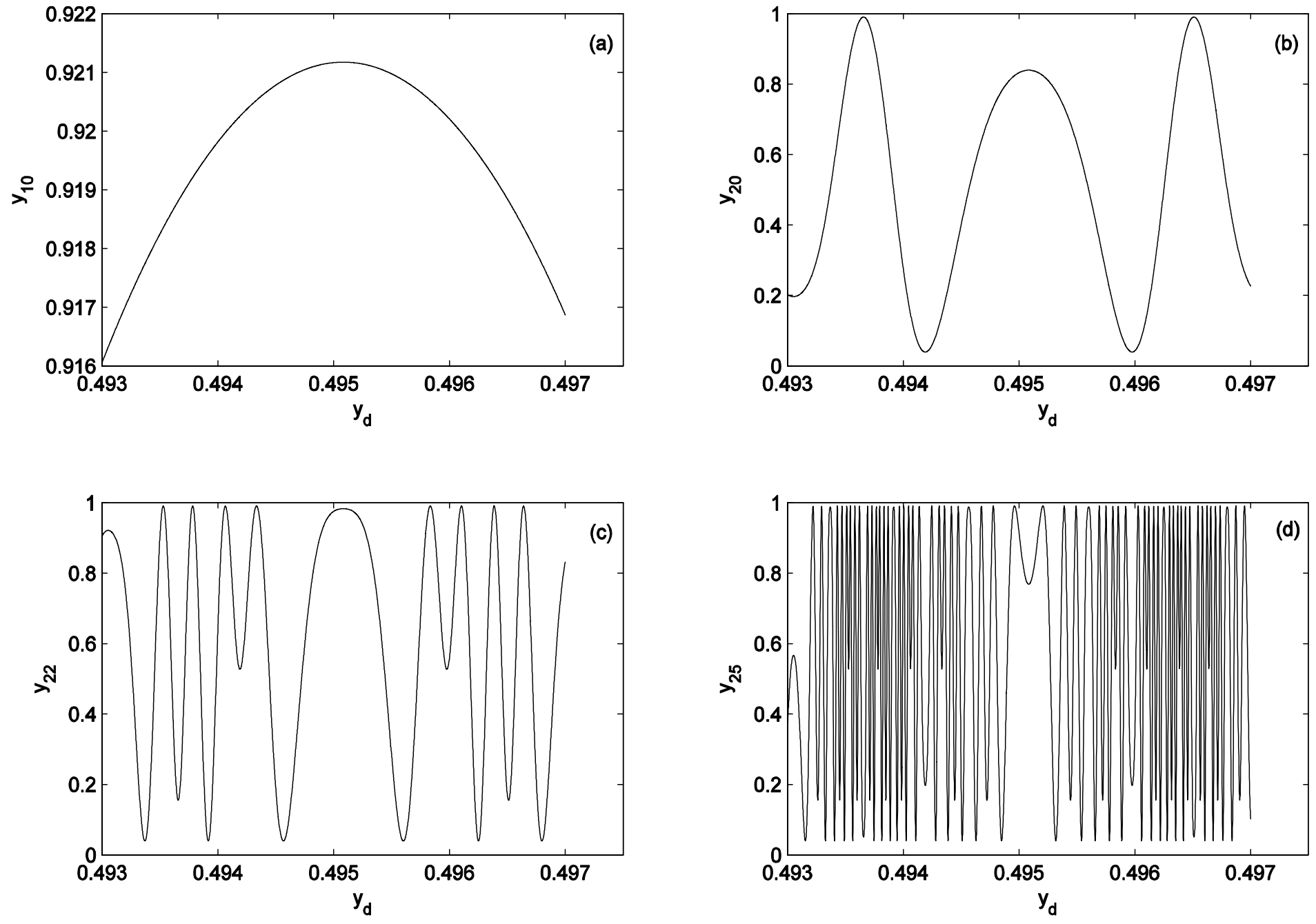

Fig. 2

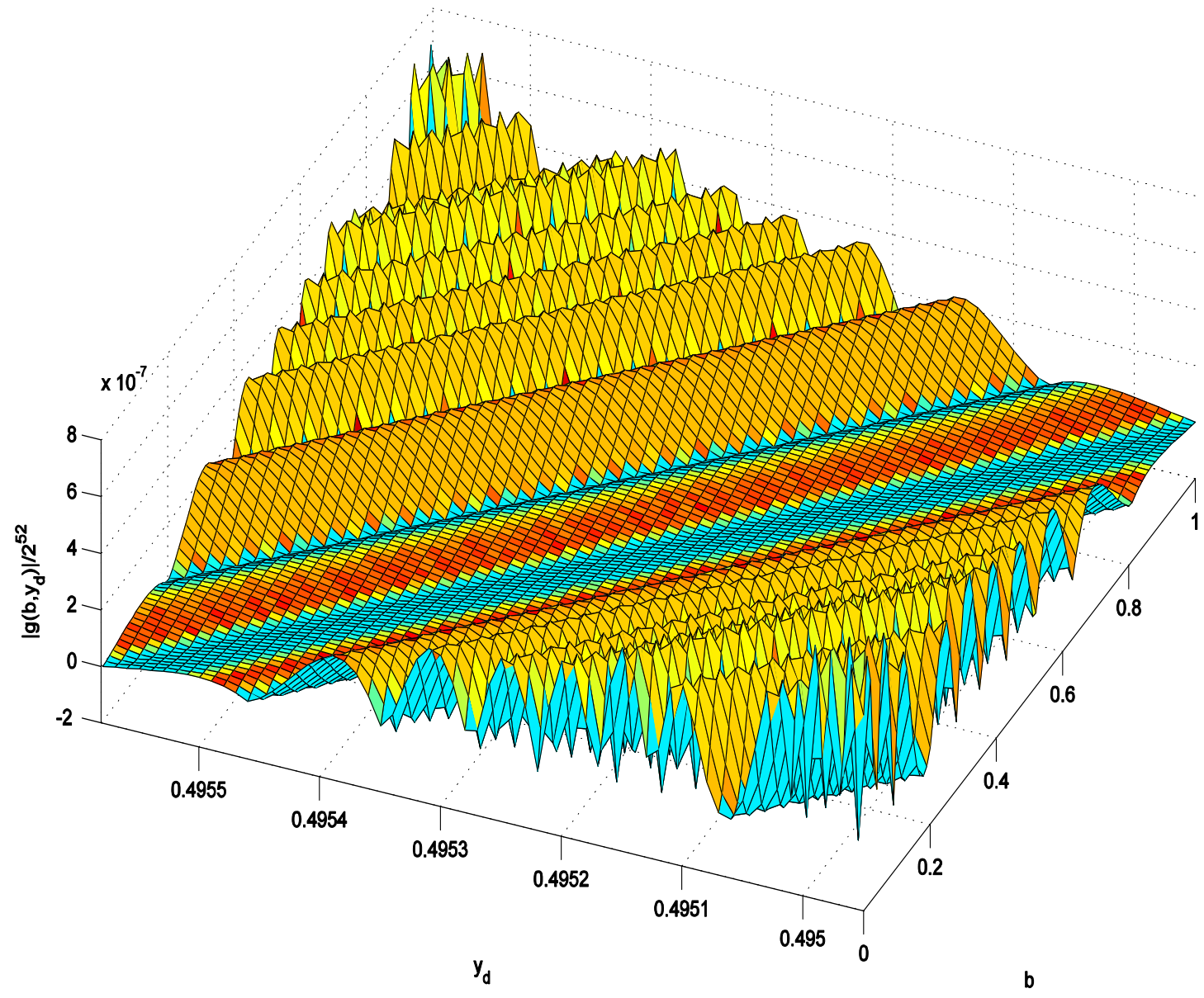

Fig. 3

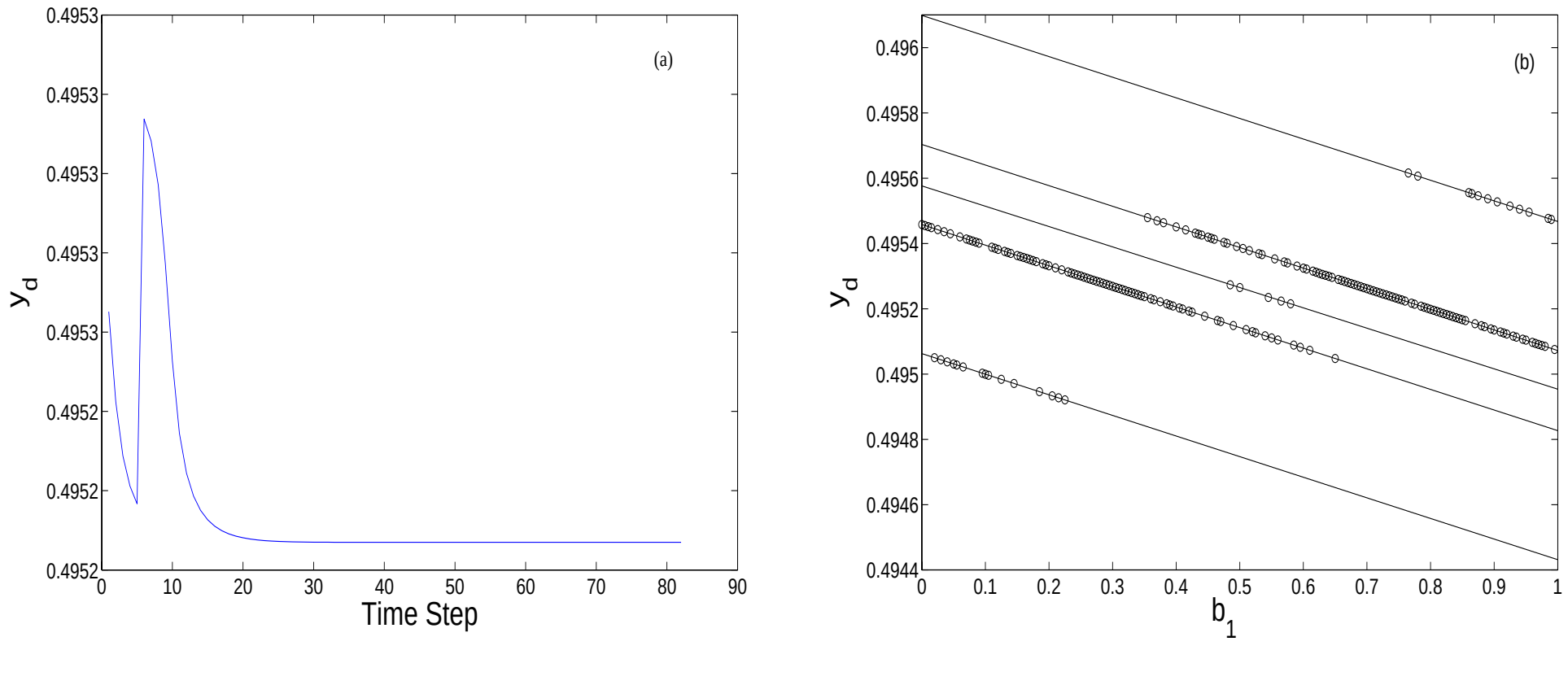

Fig.4

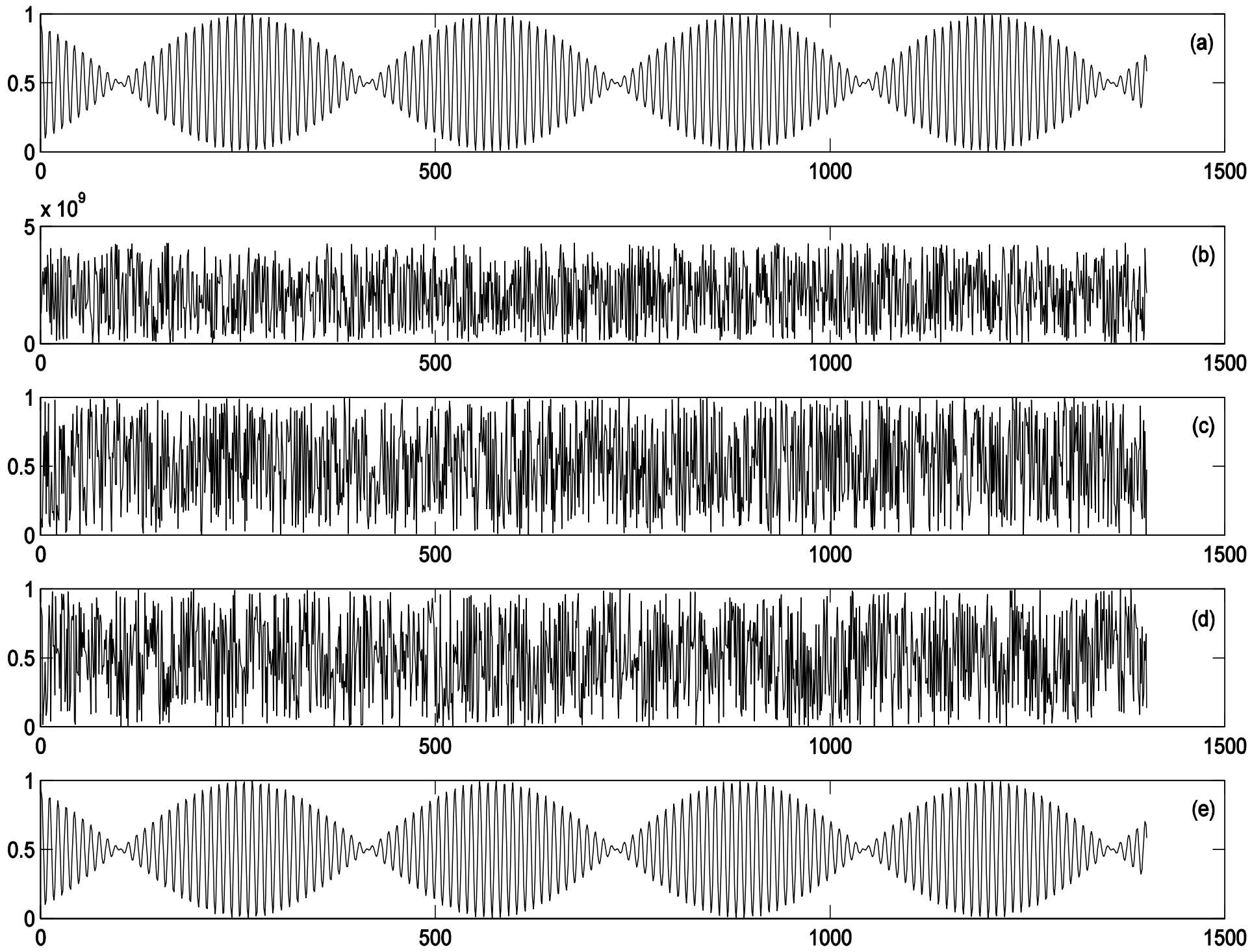

Fig. 5

Captions of Tables

TABLE 1: The equation of the lines which contain all possible $(b^*, y^*)$.

Table 1

| Line Index | $A$ | $B$ |
|---|---|---|
| 1 | -0.000631312462327782 | 0.495062818034644914 |
| 2 | -0.000631313093178722 | 0.495458106423557865 |
| 3 | -0.000622651840623326 | 0.495576481606239128 |
| 4 | -0.000631313094171622 | 0.495703510840223460 |
| 5 | -0.000631312458570371 | 0.496098818026457278 |